%% file: smashing-woot.tex
\documentclass[twocolumn,english]{article}
\usepackage{lmodern}
\usepackage[T1]{fontenc}
\usepackage[latin9]{luainputenc}
\usepackage{babel}
\usepackage{array}
\usepackage{url}
\usepackage{graphicx}
\usepackage[unicode=true,pdfusetitle,
 bookmarks=true,bookmarksnumbered=false,bookmarksopen=false,
 breaklinks=false,pdfborder={0 0 1},backref=false,colorlinks=false]
 {hyperref}

\makeatletter

\providecommand{\tabularnewline}{\\}

 \usepackage{usenix}

\g@addto@macro{\UrlBreaks}{\UrlOrds}

\@ifundefined{showcaptionsetup}{}{%
 \PassOptionsToPackage{caption=false}{subfig}}
\usepackage{subfig}
\makeatother

\begin{document}

\title{Shattered Trust: When Replacement Smartphone Components Attack}

\author{	Omer Shwartz, Amir Cohen, Asaf Shabtai, Yossi Oren\\
	Ben-Gurion University of the Negev\\
		omershv@post.bgu.ac.il, amir3@post.bgu.ac.il, shabtaia@bgu.ac.il, yos@bgu.ac.il}
\maketitle
\begin{abstract}
Phone touchscreens, and other similar hardware components such as
orientation sensors, wireless charging controllers, and NFC readers,
are often produced by third-party manufacturers and not by the phone
vendors themselves. Third-party driver source code to support these
components is integrated into the vendor\textquoteright s source code.
In contrast to ``pluggable'' drivers, such as USB or network drivers,
the component driver's source code implicitly assumes that the component
hardware is authentic and trustworthy. As a result of this trust,
very few integrity checks are performed on the communications between
the component and the device's main processor. 

In this paper, we call this trust into question, considering the fact
that touchscreens are often shattered and then replaced with aftermarket
components of questionable origin. We analyze the operation of a commonly
used touchscreen controller. We construct two standalone attacks,
based on malicious touchscreen hardware, that function as building
blocks toward a full attack: a series of \textbf{touch injection}
attacks that allow the touchscreen to impersonate the user and exfiltrate
data, and a \textbf{buffer overflow} attack that lets the attacker
execute privileged operations. Combining the two building blocks,
we present and evaluate a series of \textbf{end-to-end} attacks that
can severely compromise a stock Android phone with standard firmware.
Our results make the case for a hardware-based physical countermeasure. 
\end{abstract}

\section{Introduction}

\input{introduction.tex}

\section{\label{sec:Related-Work}Related Work}

\input{related-work.tex}

\section{\label{sec:Risk-of-malicious}Our Attack Model}

\input{hardware-attacks.tex}

\section{\label{sec:Touchscreen-Hardware}Reverse Engineering a Touch Screen}

\input{touchscreen-hardware.tex}

\section{\label{sec:Attack-Building-Blocks}Attack Building Blocks}

\input{attacking-the-phone.tex}

\section{\label{sec:End-to-End-Attacks}End-to-End Attacks}

\input{end-to-end-attack.tex}

\section{\label{sec:Attacking-Additional-Devices}Attacking Additional Devices}

\input{additional-devices.tex}

\section{Discussion}

\input{discussion.tex}

\subsection*{Acknowledgements}

This research was supported by Israel Science Foundation grants 702/16
and 703/16.

\bibliographystyle{unsrt}
\bibliography{smashing}

\newpage{}

\begin{figure*}
\begin{centering}
\includegraphics[width=0.8\textwidth]{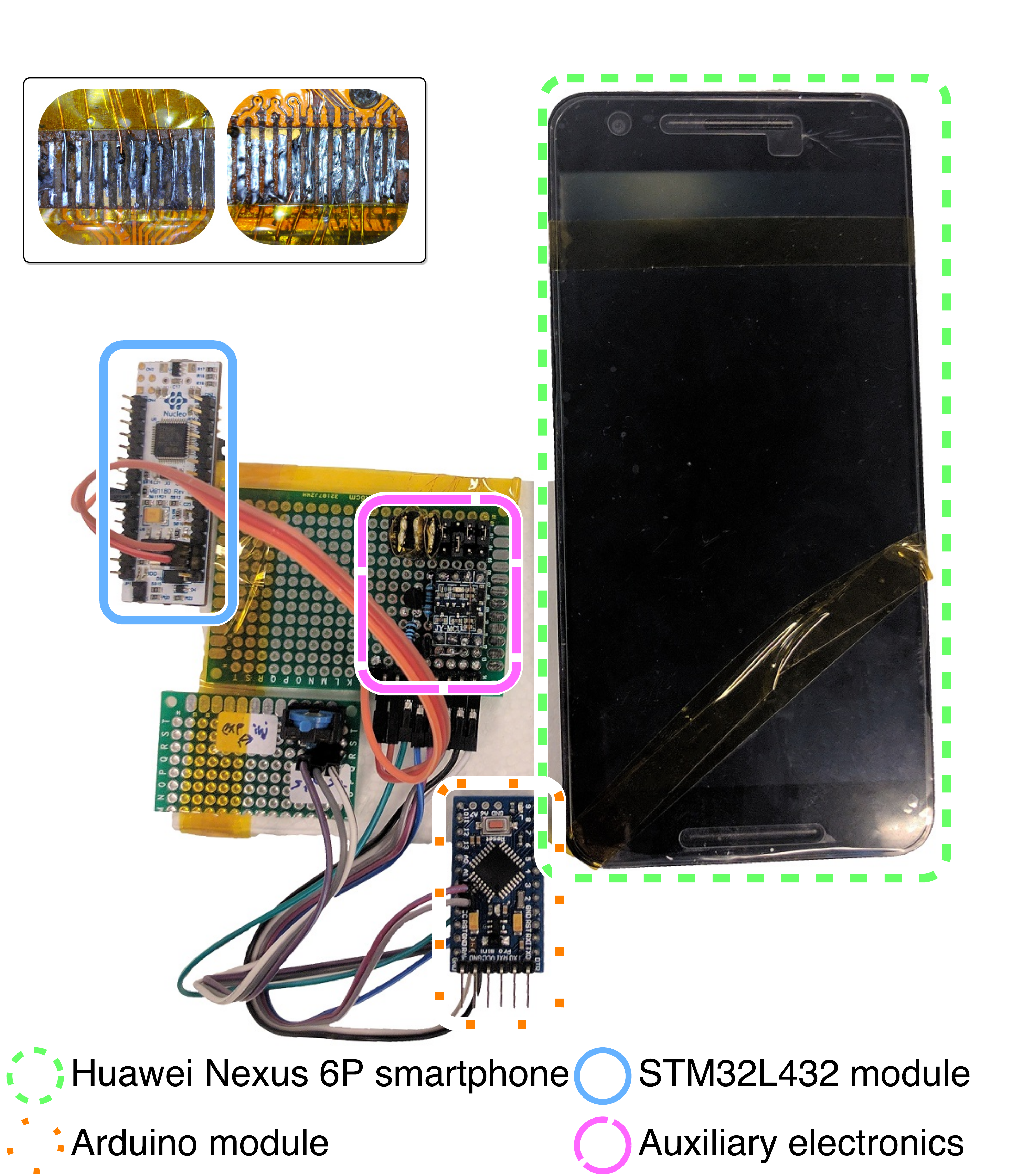}
\par\end{centering}
\caption{\label{fig:attack-setup}The complete attack setup. The figure shows
an exposed touch controller interface wired to a prototyping board
embedded with auxiliary electronics and connected to an Arduino micro-controller
module. The prototyping board is also connected to an STM32L432 micro-controller
module which is used for debugging purposes. Inset: wires soldered
onto the touch controller communication connection}
\end{figure*}

\end{document}

%% file: introduction.tex
Mobile phones are often dropped, shattering their screens. According
to a 2015 study, more than 50\% of global smartphone owners have damaged
their phone screen at least once, and 21\% of global smartphone owners
are currently using a phone with a cracked or shattered screen \cite{cracked}.
While phones suffering from fractured screens may be repaired at phone
vendor-operated facilities such as Apple Stores, it is often more
convenient and cost-effective for phone users to use third-party repair
shops. Some technically savvy users may even purchase touchscreen
replacement kits from online marketplaces such as eBay and perform
the repair themselves. These types of unofficial repairs tend to include
the cheapest possible components, and thus may introduce, knowingly
or unknowingly, counterfeit or unbranded components into the phone.

\begin{figure}
\begin{centering}
\includegraphics[width=0.6\columnwidth]{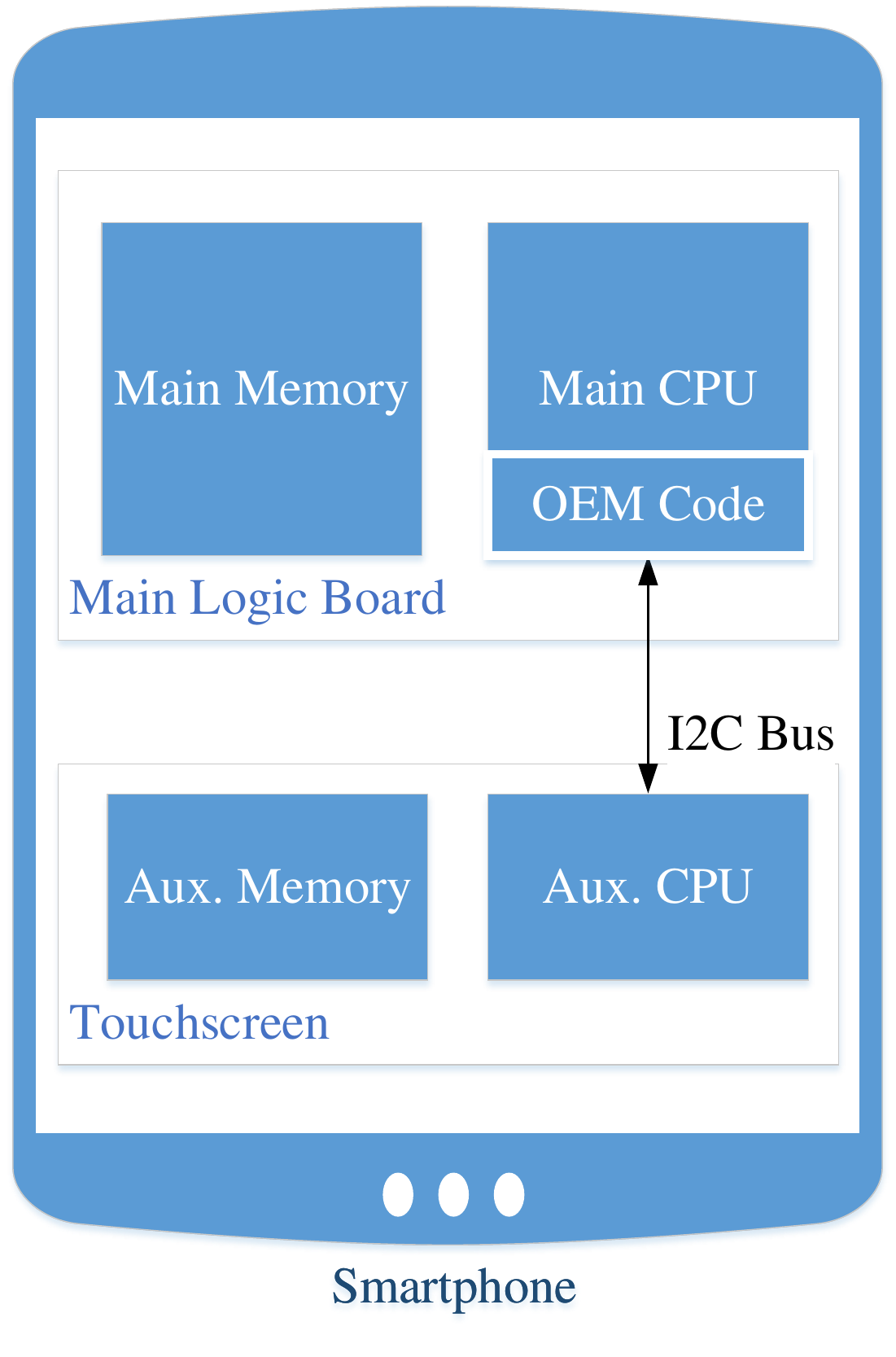}
\par\end{centering}
\caption{\label{fig:phone-touchscreen-driver-interface}The smartphone, its
touch screen, and its associated device driver software.}
\end{figure}

Phone touchscreens, and other similar hardware components such as
orientation sensors, wireless charging controllers, and near-field
communications (NFC) readers, are seldom produced by the phone vendors
themselves. Instead, original equipment manufacturers (OEMs) such
as Synaptics, MediaTek and Maxim Integrated provide these components,
as well as the device driver source code, to phone vendors who integrate
the components into their phones. The vendors then proceed to integrate
this code into their own source code, making minor adjustments to
account for minor differences between device models, such as memory
locations, I/O bus identifiers, etc. These minor tweaks and modifications
make the process of creating and deploying patches for such device
drivers a very difficult endeavor, as we discuss further in Section
\ref{sec:Related-Work}. The example in Figure \ref{fig:phone-touchscreen-driver-interface}
illustrates this setting. The smartphone's main logic board runs a
specific OEM code (the device driver) that communicates with the touchscreen
over the internal bus using a simple, common interface. Even hardened,
secure, or encrypted phones, such as those used by governmental and
law enforcement agencies, often use commercial operating systems and
a driver architecture that follows the same paradigm \cite{disaApproved}.

The key insight of this paper starts with the observation that the
device drivers (written by the OEMs and slightly modified by the phone
vendors) exist \emph{inside} the phone's trust boundary. In contrast
with drivers for ``pluggable'' peripherals such as USB accessories,
these OEM drivers assume that the internal components they communicate
with are also inside the phone's trust boundary. However, we observe
that these internal components are quite emphatically \emph{outside}
the smartphone's trust boundary. Indeed, there are some hundreds of
millions of smartphones with untrusted replacement screens. Our research
question was therefore: How might a malicious replacement peripheral
abuse the trust given to it by its device driver? How can we defend
against this form of abuse?

Hardware replacement is traditionally considered a strong attack model,
under which almost any attack is possible. Uniquely in our case, we
add an important restriction to this model: we assume only a specific
component, with an extremely limited hardware interface, is malicious.
Furthermore, we assume that the repair technician installing the component
is uninvolved. Hundreds of millions of devices satisfying this assumption
exist in the wild. One can assume that these limitations make this
attack vector weaker than complete hardware replacement; we show that
it is not.

In this work we highlight the risk of counterfeit or malicious components
in the consumer setting, where the target is the user's privacy, personal
assets, and trust. We show how a malicious touchscreen can record
user activity, take control of the phone and install apps, direct
the user into phishing websites and exploit vulnerabilities in the
operating system kernel in order to gain privileged control over the
device. Since the attack is carried out by malicious code running
out of the CPU's main code space, the result is a \emph{fileless attack,
}which cannot be detected by anti-virus software, leaves no lasting
footprint and surviving firmware updates and factory resets.

Our paper makes the following contributions:
\begin{enumerate}
\item We survey the risk of malicious peripheral attacks on consumer devices
and argue that this avenue of attack is both practical and effective. 
\item We introduce the design and architecture of touchscreen assemblies
and touch controllers, along with their communication protocols, limiting
our scope to smartphones and their screens. In addition, we analyze
the operation of a commonly used touch controllers (Synaptics S3718)
and their communications with the device driver.
\item We describe two attack building blocks that can be used in a larger
attack: a \textbf{touch injection} attack that allows the touchscreen
to impersonate the user, and a \textbf{buffer overflow} attack that
lets the attacker execute privileged operations.
\item Combining the two building blocks, we present a series of \textbf{end-to-end}
attacks that can severely compromise a stock Android phone with standard
firmware. We implement and evaluate three different attacks, using
an experimental setup based on a low-cost micro-controller embedded
in-line with the touch controller communication bus. These attacks
can:
\begin{enumerate}
\item \textbf{Impersonate the user} - By injecting touch events into the
communication bus, an attacker can perform any activity representing
the user. This includes installing software, granting permissions
and modifying the device configuration. 
\item \textbf{Compromise the user} - An attacker can log touch events related
to sensitive operations such as lock screen patterns, credentials
or passwords. An attacker can also cause the phone to enter a different
state than the one expected by the user by injecting touch events.
For example, we show an attack that waits until a user enters a URL
for a website and then stealthily modifies the touches to enter a
URL of a phishing website, thereby causing the user surrender his
or her private information. 
\item \textbf{Compromise the phone} - By sending crafted data to the phone
over the touch controller interface, an attacker can exploit vulnerabilities
within the device driver and gain kernel execution capabilities. 
\end{enumerate}
\item To demonstrate the generality of our attack method, we show how we
ported our attack to another device (Atmel T641) using similar techniques
and tools. 
\end{enumerate}

%% file: related-work.tex
Throughout the relatively short history of smartphones, both malware
and protection mechanisms have evolved drastically to fit this emerging
platform. Android malware in particular has been shown to utilize
privilege escalation, siphon private user data and enlist phones into
botnets \cite{zhou2012dissecting}. Bickford et al. \cite{bickford2010rootkits}
address the topic of smartphone rootkits, defining a rootkit as ``a
toolkit of techniques developed by attackers to conceal the presence
of malicious software on a compromised system''. Malicious activities
performed by rootkits include wiretapping into phone calls, exfiltration
of positioning information and battery exhaustion denial of service.

Hardware interfaces have recently been a subject of concern for security
researchers in the personal computer setting, due to their involvement
in highly privileged processes \cite{corbet2005linux}. Hardware components
enjoying Direct Memory Access (DMA) such as the Graphics Processing
Unit (GPU) can implant malware within the kernel memory \cite{danisevskis2013dark}.
Ladakis el al. \cite{ladakis2013you} demonstrate a GPU based keylogger
where the GPU abuses its DMA capabilities for monitoring the keyboard
buffer inside the kernel memory and saving keystroke information to
the GPU memory. Brocker et al. \cite{brocker2014iseeyou} used the
firmware update functionality of a MacBook iSight camera for installing
malicious firmware on the camera. Using their firmware, the authors
show the ability of taking discrete photos or videos without turning
on the indicator light that informs the user about the usage of the
camera. Additionally, the authors use their firmware for enumerating
the camera as a USB keyboard and present the ability of the device
to escape virtual machine confinement. 

Most of the existing works dealing with hardware interfaces focus
on hardware components that can either be updated by the user or easily
replaced. Smartphones are more monolithic by design than PCs, their
hardware inventory is static and components can only replaced with
matching substitutes. The smartphone operating system contains device
firmwares that can only be updated alongside the operating system.
Thus, there is far less of a research focus on smartphone hardware,
based on the assumption that it cannot be easily replaced or updated
and is therefore less exposed to the threats discussed above. We challenge
this assumption, noting that smartphone components are actually being
replaced quite frequently and often with non genuine parts, as we
show in Section \ref{sec:Risk-of-malicious}. 

The troubles that may come with counterfeit components had not been
completely ignored by the mobile industry. An example is the ``error
53'' issue experienced by some iPhone users after replacing their
fingerprint sensors with off-brand ones and failing validity checks
\cite{appleError53}. However, it seems like these kind of validity
checks are not widely accepted, since counterfeit replacements usually
pass unnoticed. The risk of counterfeit components had also been raised
in the national security setting in a National Institute of Standards
(NIST) draft, putting emphasis on supply chains \cite{nistDraft}.

Zhou et al. \cite{zhou2014peril} performed a systematic study of
the security hazards in Android device customizations. The authors
designed a tool, ADDICTED, that detects customization hazards. The
authors raised the concern that the customizations performed by vendors
can potentially undermine Android security. In a previous work \cite{ourPaper}
we focused on driver customizations, reviewing the source code of
26 Android phones and mapping the device drivers that are embedded
in the kernel of their operating system. Our survey found a great
deal of diversity in OEMs and device drivers. Each phone contained
different driver software, and there were few common device drivers
between the tested phones. This landscape makes it difficult to patch,
test and distribute fixes for vulnerabilities found in driver code.

%% file: hardware-attacks.tex
Counterfeit components have been in existence ever since the dawn
of the industrial age. Their effectiveness as attack vectors is also
well known. What, then, is unique about the particular setting of
a smartphone? We argue that our specific attack model is a unique
restriction the hardware replacement attack model: we assume only
a specific component, with an extremely limited hardware interface,
is malicious, while the rest of the phone (both hardware and software)
can still be trusted. Furthermore, we assume that the repair technician
installing the component is not malicious, and will perform no additional
operations other than replacing the original component with a malicious
one. Hundreds of millions of devices satisfying this attack model
exist in the wild. One can assume that these limitations make this
attack vector weaker than complete hardware replacement; we show that
it is not. On the contrary, the nature of the smartphone ecosystem
makes this attack model both \emph{practical} and \emph{effective}.

\begin{figure*}
\begin{centering}
\includegraphics[width=0.8\textwidth]{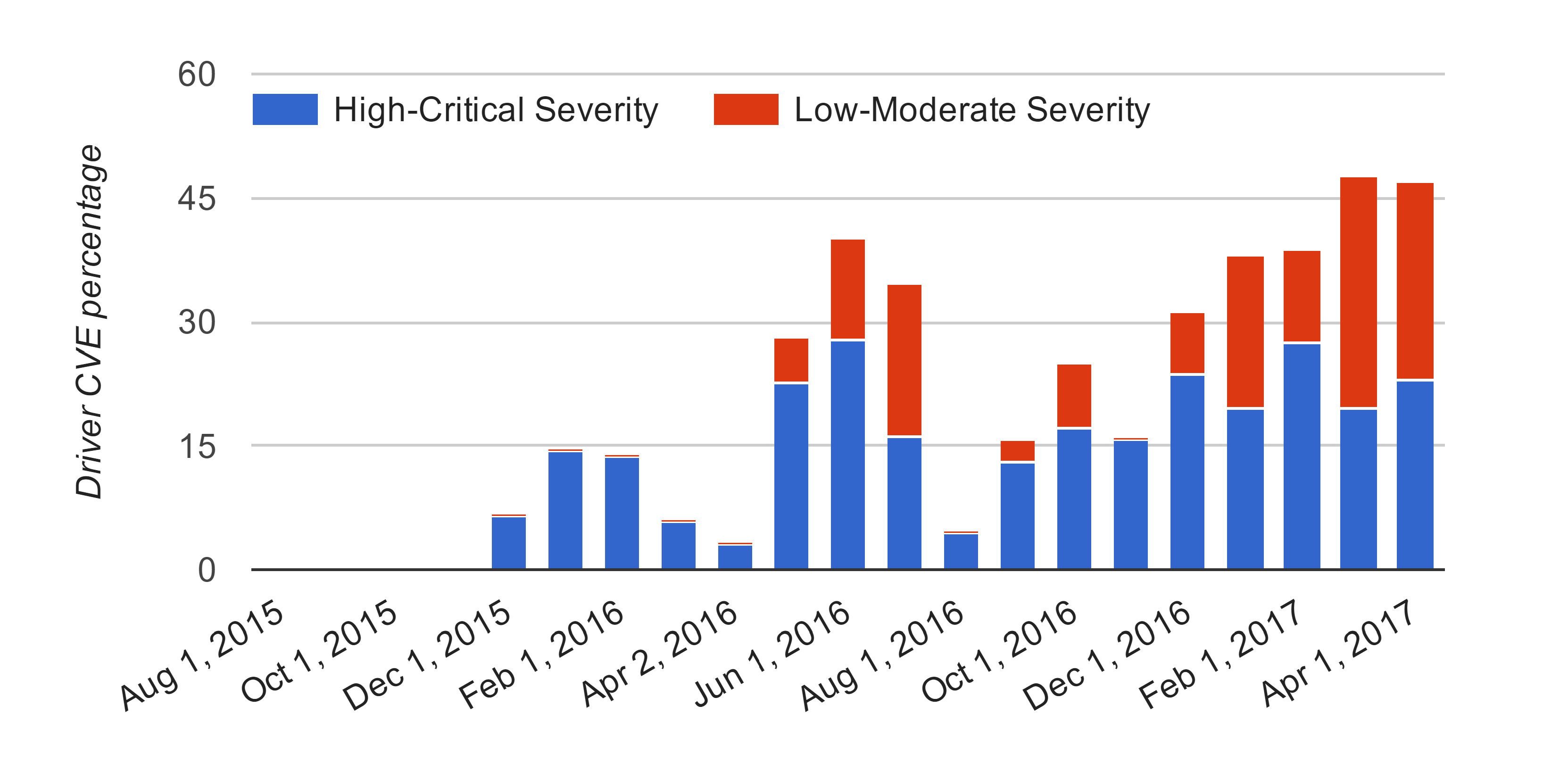}
\par\end{centering}
\caption{\label{fig:The-ratio-of}The ratio of patched Android CVEs that occur
in drivers out of all patched Android CVEs. The figure was compiled
using information from the Android security bulletin \cite{secbull}.}
\end{figure*}

The pervasiveness of untrusted components in the smartphone supply
chain was investigated in September 2016 by Underwriters Laboratories
\cite{Counterfeit}. UL researchers obtained 400 iPhone charging adapters
from multiple sources in eight different countries around the world,
including the U.S., Canada, Colombia, China, Thailand and Australia,
and discovered that \emph{nearly all of them} were counterfeited and
contained sub-standard hardware. Similarly, in October 2016 Apple
filed a lawsuit against Amazon.com supplier Mobile Star LLC, claiming
that \textquotedblleft Apple {[}...{]} has purchased well over 100
iPhone devices, Apple power products, and Lightning cables sold as
genuine by sellers on Amazon.com {[}...and{]} revealed \emph{almost
90\% of these products are counterfeit}.''\cite{CounterfeitApple}.
Considering the condition of the third-party marketplace, one can
assume with high confidence that unless a phone has been repaired
at a vendor-operated facility such as an Apple Store, it is likely
to contain counterfeit components.

Conservative estimates assume that there are about 2 billion smartphones
in circulation today. Assuming that 20\% of these smartphones have
undergone screen replacement \cite{cracked}, there are on the order
of 400 million smartphones with replacement screens in the world.
An attack which compromises even a small fraction of these smartphones
through malicious components will have a rank comparable to that of
the largest PC-based botnets.

Let us next assume that a malicious peripheral, such as a touchscreen,
has made it into a victim's smartphone. What sort of damage can it
cause?

As stated in \cite{ourPaper}, attacks based on malicious hardware
can be divided into two different classes. \emph{First-order attacks
}use the standard interaction modes of the component, but do so without
the user\textquoteright s knowledge or consent. In the specific case
of a malicious touchscreen, the malicious peripheral may log the user\textquoteright s
touch activity or impersonate user touch events in order to impersonate
the user for malicious purposes. We demonstrate some of these attacks
in Subsection \ref{subsec:Touch-Logging-and} \emph{Second order attacks}
go beyond exchanging properly-formed data, and attempt to cause a
malfunction in the device driver and compromise the operating system
kernel. Such an attack requires that the peripheral send malformed
data to the CPU, causing the device driver to malfunction and thereby
compromising the operating system kernel. Once the kernel is compromised,
it is possible to disable detection and prevention of suspicious system
activity, eavesdrop on sensors and on other applications, and most
significantly operate on systems where only a partial software stack
had been loaded, such as a device in charging, standby or even turned
off state\cite{butt2009protecting,okolie2013penetration,zhou2014peril,guri2015joker,suarez2014evolution}. 

While \emph{first order attacks} require no software vulnerability
and can be performed by any peripheral contained in consumer electronics,
\emph{second order attacks }require a vulnerability to be exploited.
The ability of malicious \emph{pluggable} peripherals to compromise
the smartphone is very well demonstrated. A review of 1077 Android
CVEs (Common Vulnerabilities and Exposures) patched between August
2015 and April 2017 shows that at least 29.5\% (318 items) take place
in the device driver context \cite{secbull}. Figure \ref{fig:The-ratio-of}
shows the growth in driver related CVEs. The fact that device driver
vulnerabilities are often detected in the pluggable setting, combined
with a general lack of attention to the internal component setting,
indicated that to us that it was very likely that internal components
might be used to trigger vulnerabilities just like pluggable components.
In this paper we describe two such vulnerabilities we found in common
touchscreen drivers (Synaptics S3718 and Atmel T641).

%% file: touchscreen-hardware.tex
Even though touchscreen assemblies have different functions, capabilities
and physical properties according to the phone model that houses them,
most of these assemblies have a similar general design. In this Subsection
we introduce the key components of the touchscreen assembly and their
functions with a focus on the workings of the Huawei Nexus 6P smartphone
touchscreen assembly containing the Synaptics S3718 touch controller.
In Subsection \ref{sec:Attacking-Additional-Devices} we extend our
analysis to another phone model.

Information regarding the Nexus 6P's touchscreen functionality was
obtained by reviewing the open source code for the Synaptics device
driver available in the Google MSM kernel repository \cite{msmkernel}
and by physical disassembly of a phone, followed by reverse engineering
the communication protocol using a Saleae logic analyzer.

\subsection{Touchscreen Assembly}

\begin{figure*}
\begin{centering}
\subfloat[\label{fig:TS-assembly}The Nexus 6P touchscreen assembly flex printed
circuit board, located on the back side of the touchscreen.]{

\includegraphics[width=0.45\textwidth]{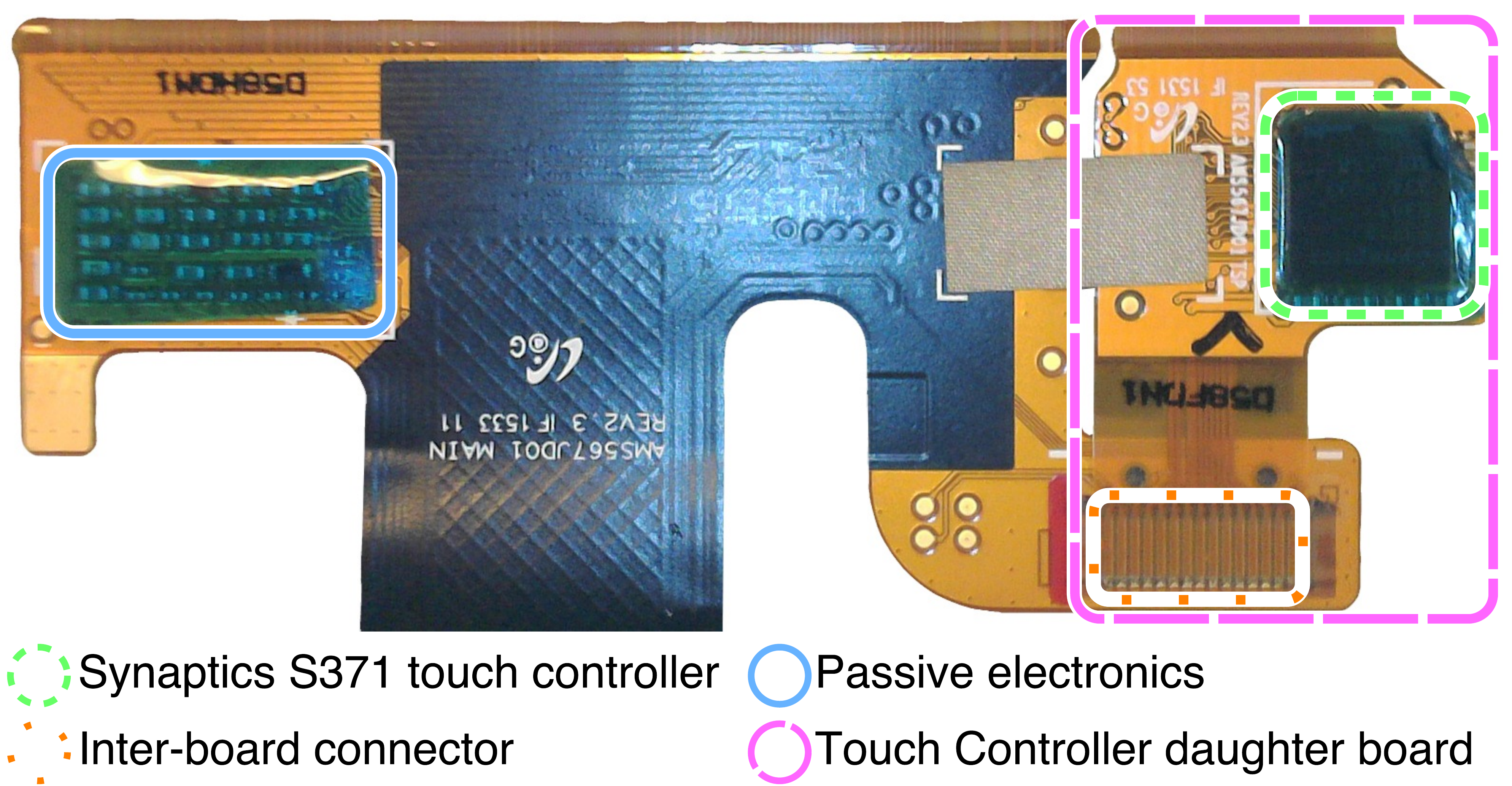}}\hfill{}\subfloat[\label{fig:TS connector}The connection between the touch controller
daughter board and the main touchscreen assembly daughter board of
an assembly for the Nexus 6P. Marked: relevant pinout for the communication
bus.]{

\includegraphics[width=0.45\textwidth]{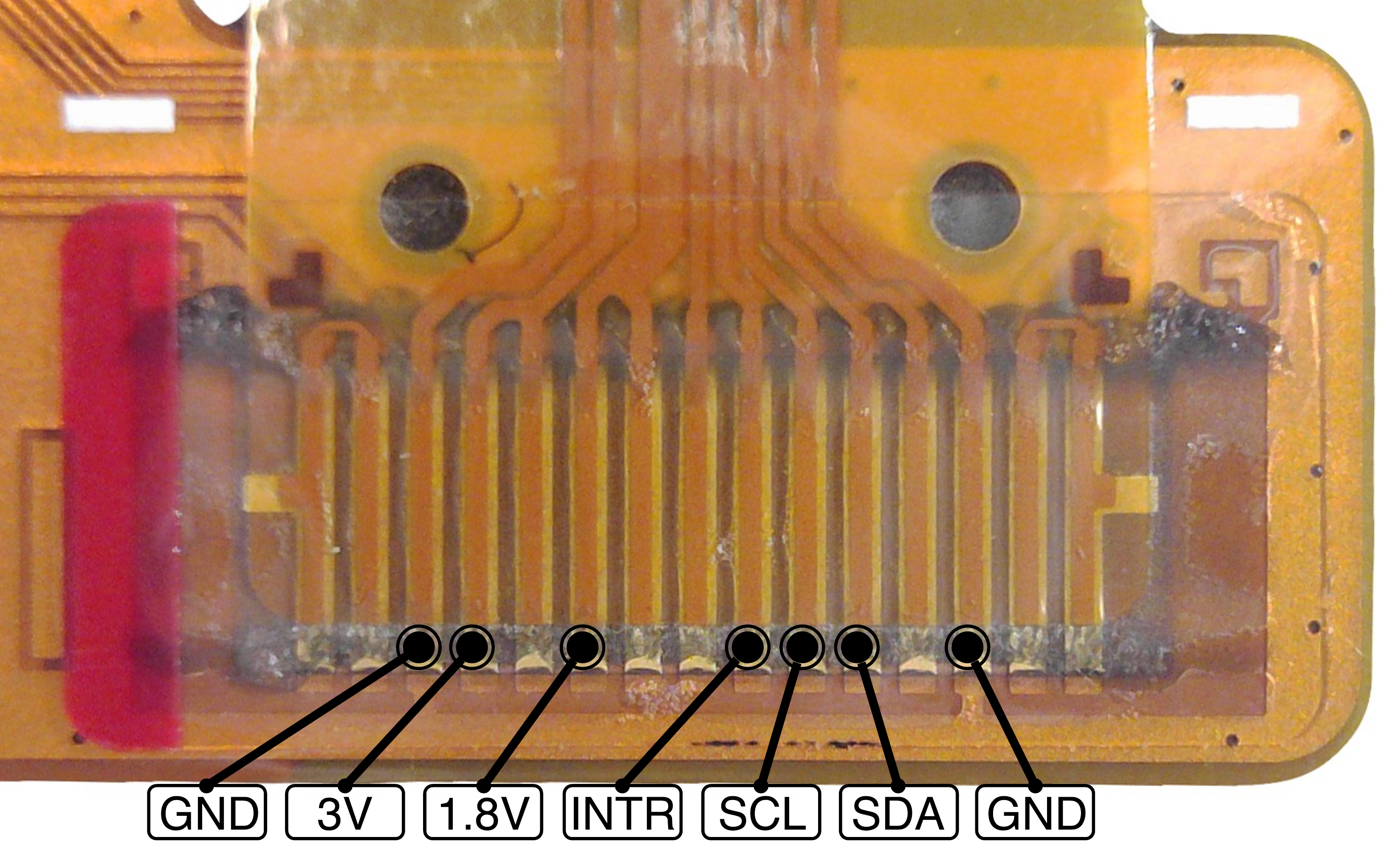}}
\par\end{centering}
\caption{}
\end{figure*}

A touchscreen assembly most essentially contains a \textbf{display
device}, such as an Liquid Crystal Display (LCD) or an Organic Light-Emitting
Diode Display (OLED). Layered on top of the display is a thin, transparent
capacitive or resistive \textbf{sensing surface} allowing accurate
positioning of physical events. The sensing functionality is managed
by a \textbf{touch controller}, an integrated circuit (IC) responsible
for analyzing the signals generated by the sensing surface and translating
them into digital data. The touch controller typicaly resides on a
\textbf{daughter printed circuit board} (PCB), together with other
ICs responsible for other display-related tasks. The daughter board
also includes a connector to the main phone board. In many cases,
including the Nexus 6P touchscreen assembly, there are multiple daughter
boards, one of which is entirely dedicated to the touch capabilities.

\subsection{Touch Controller Communications}

In most smartphones, the touch controller communicates with the device
driver residing on the host processor via a dedicated Inter Integrated
Circuit (I$^{2}$C) bus \cite{NXP2014i2c}, a general purpose, low
speed bus designed for cost effective communication between Integrated
Circuits (ICs). The I$^{2}$C bus behaves as a physical layer between
master and slave devices, where master devices are allowed to read
and write from and to registers in the slave device's memory. By manipulating
these registers, the device driver (acting as master) can control
the behavior of the touch controller (acting as slave); by populating
the appropriate registers and triggering an interrupt, the touch controller
can send events to the device driver. On top of this low-level communication
interface, the device driver typically defines a proprietary layer
required for the instrumentation and operation of the touch controller.

In the Nexus 6P phone, the Synaptics S3718 touch controller daughter
board has I$^{2}$C connections on contacts SCL and SDA as seen in
Sub-Figure \ref{fig:TS connector}. It has an additional contact for
generating an interrupt notifying the host processor of touch-related
events. The I$^{2}$C bus communicates at the rate of 400 Kbps.

A basic mapping of the shared touch controller registers and functions
was extracted from the open source device driver made available by
Google. Additional reverse engineering and observation provided a
fuller picture of the protocol.

\subsubsection{Boot up process}

\begin{table*}
\centering{}\caption{\label{tab:TS functions}Partial index of the main Synaptics S3718
functions and their purpose }
\begin{tabular}{|>{\centering}p{0.09\linewidth}|>{\centering}p{0.09\linewidth}|>{\centering}p{0.1\linewidth}|>{\centering}p{0.09\linewidth}|>{\centering}p{0.09\linewidth}|>{\centering}p{0.35\linewidth}|}
\hline 
Function ID & Query Address & Command Address & Control Address & Database Address & Function Purpose\tabularnewline
\hline 
\hline 
0x01 & 0x3F & 0x36 & 0x14 & 0x06 & General control and status of the touch controller\tabularnewline
\hline 
0x12 & 0x5C & 0x00 & 0x1B & 0x08 & Reporting of simple touch events, including multi-finger touches\tabularnewline
\hline 
0x51 & 0x04 & 0x00 & 0x00 & 0x00 & Firmware update interface\tabularnewline
\hline 
\end{tabular}
\end{table*}

During the boot up process, the device driver probes the touch controller
memory and learns which functions the controller possesses. A controller
function or capability is reported through a 6 byte function descriptor.
The function descriptor contains four register addresses used for
manipulating the function along with an interrupt count that signifies
the number of interrupt types the function generates. A map of several
controller functions can be seen in Table \ref{tab:TS functions}.
After probing and querying for the functions, the device driver checks
the installed firmware against the firmware file embedded in the kernel
memory and triggers a firmware update if necessary. Eventually, the
device driver enables the appropriate handlers for all function specific
interrupts and writes the configuration data to the relevant functions.

\subsubsection{Touch reporting}

In order to generate a touch event, the touch controller electrically
pulls the interrupt line towards the ground and thus notifies the
device driver of an incoming event. The device driver in turn reads
the interrupt register 0x06 and deduces which of the touch controller
functions generated the interrupt. In the case of a normal touch event
this will be function 0x12. The device driver continues to read a
bitmap of the fingers involved in this event from register 0x0C and
eventually reads register 0x08 for a full inventory of the touch event.

%% file: attacking-the-phone.tex
This section describes two basic attacks that severely compromise
the phone. The first attack allows the attacker to record, intercept,
and inject touch events. The second attack leverages vulnerabilities
discovered in the operating system kernel and executes privileged
arbitrary code. Our attack assumes that the phone's touch controller
had been replaced with a malicious component, but that the rest of
the hardware and software on the phone is authentic and trusted.

\subsection{\label{subsec:Attack-Setup}Attack Setup}

The attacks were demonstrated on a Huawei Nexus 6P smartphone running
the Android 6.0.1 operating system, build MTC19X. The phone is operating
with stock manufacturer firmware and has been restored to factory
state with a memory wipe using the ``factory data reset'' feature
in the settings menu.

The touch screen assembly was separated from the rest of the phone
and the touch controller daughter board was located, as seen in Figure
\ref{fig:TS-assembly}. Using a hot air blower on the connection between
the touch controller daughter board and the main assembly daughter
board we were able to separate the boards and access the copper pads.
The copper pads were soldered to thin enameled copper wire that was
attached to a prototyping board. Using this setup, we were able to
simulate a chip-in-the-middle scenario in which a benign touchscreen
has been embedded with a malicious integrated chip that manipulates
the communication bus. A high-resolution image of our attack setup
can be found in the Appendix.

Our attack used a commonly available Arduino platform \cite{arduino}
based on the ATmega328 micro-controller for our attack. A setup such
as the one described above can easily be minimized in a factory or
a skilled shop in order to fit on the touchscreen assembly daughter
board. ATmega328, the programmable micro-controller used in our attacks,
is available in packages as small as 4 x 4 x 1 mm \cite{atmega}.
Other, more powerful micro-controllers are available in smaller footprints
of 1.47 x 1.58 x 0.4 mm and less \cite{cypress}. Since the data sent
by our attack fully conforms to the layer 2 and layer 1 parts of the
I$^{2}$C specification, it can also be implemented in the firmware
of the malicious peripheral's internal micro-controller, removing
the need for additional hardware altogether.

\subsection{Touch Logging and Touch Injection\label{subsec:Touch-Logging-and}}

In this attack, the malicious micro-controller eavesdrops on user
touch events (touch logging) and injects generated touch events into
the communication bus (touch injection). The micro-controller software
behind the phishing attack is built of three components: two state
machines, one maintaining a keyboard mode and the other maintaining
a typing state; and a database that maps screen regions to virtual
keyboard buttons. The state machine holding the keyboard modes changes
state when a keyboard mode switch key had been pressed. The basic
Nexus 6P keyboard has four modes: English characters, symbols, numbers,
and emoji. The typing state machine is used for tracking the typed
characters and matching them to specified trigger events (such as
typing in a URL). Complex context information, such as keyboard orientation,
language and activity, has been shown to be detectable from low-level
touch events by other authors \cite{DBLP:journals/tifs/FrankBMMS13}.
When the required trigger event is reached, touch injection is triggered
and a set of generated touch events is sent on the communication line.
Our current hardware is capable of creating touch events at a rate
of approximately 60 taps per second.

\subsection{Arbitrary Code Execution \label{subsec:Arbitrary-Code-Execution}}

This attack exploits vulnerabilities in the touch controller device
driver embedded within the operating system kernel in order to gain
arbitrary code execution within the privileged kernel context. A chain
of logical manipulations on performed by the malicious micro-controller
causes a heap overflow in the device driver that is further exploited
to perform a buffer overflow.

\subsubsection{Design}

As a part of the boot procedure, the device driver queries the functionality
of the touch controller. We discovered that by crafting additional
functionality information we can cause the device driver to discover
more interrupts than its internal data structure can contain, causing
a heap overflow. Using the heap overflow we were able to further increase
the amount of available interrupts by overrunning the integer holding
that value. Next, an interrupt was triggered causing the device driver
to request an irregularly-large amount of data and cause a buffer
overflow. The buffer overflow was exploited using a Return Oriented
Programming (ROP) \cite{hund2009return} chain designed for the ARM64
architecture.

\subsubsection{Implementation}

\begin{table*}
\centering{}\caption{\label{tab:ROP-chain}ROP chain designed for the ARM64 architecture.
This chain results in a call to a predefined function with two arguments.}
\small%
\begin{tabular}{|>{\centering}p{0.07\textwidth}|>{\centering}p{0.5\textwidth}|>{\centering}p{0.32\textwidth}|}
\hline 
Gadget Order & Gadget Code & Relevant Pseudocode\tabularnewline
\hline 
\hline 
1 & ldp x19, x20, {[}sp, \#0x10{]} ; ldp x29, x30, {[}sp{]}, \#0x20 ;
ret; & Load arguments from stack to registers X19 and X20\tabularnewline
\hline 
2 & mov x2, x19 ; mov x0, x2 ; ldp x19, x20, {[}sp, \#0x10{]} ; ldp x29,
x30, {[}sp{]}, \#0x30 ; ret; & Assign X2 := X19; Load arguments from stack to registers X19 and X20\tabularnewline
\hline 
3 & mov x0, x19 ; mov x1, x20 ; blr x2 ; ldp x19, x20, {[}sp, \#0x10{]}
; ldr x21, {[}sp, \#0x20{]} ; ldp x29, x30, {[}sp{]}, \#0x30 ; ret; & Assign X0 := X19; Assign X1 := X20; Call X2(X0, X1)\tabularnewline
\hline 
\end{tabular}\normalsize
\end{table*}

When triggered, the malicious micro-controller shuts down power to
the touch controller and begins imitating normal touch controller
behavior. During boot, the malicious micro-controller emulates the
memory register image of the touch controller and responds in the
same way to register writes using a state machine. When probed for
function descriptors in addresses higher than 0x500 that normally
do not exist within the touch controller, the micro-controller responds
with a set of crafted function descriptors designed to cause the interrupt
register map to exceed its boundaries. Within the device driver, a
loop iterates over the interrupt register map and writes values outside
the bounds of an interrupt enable map, causing the integer holding
the number of interrupt sources to be overwritten. After waiting 20
seconds for the boot procedure to complete, the micro-controller initiates
an interrupt by pulling the interrupt line towards the ground. The
device driver which should then read up to four interrupt registers,
instead reads 210 bytes, causing a buffer overflow. Within the 210
bytes requested from the touch controller that are sent reside a ROP
chain that calls the Linux kernel function mem\_text\_write\_kernel\_word()
that writes over protected kernel memory with a chosen payload. Table
\ref{tab:ROP-chain} contains additional information about the ROP
chain.

\subsubsection{Evaluation}

Four different payloads were mounted on top of the ROP chain described
above and tested in attack scenarios on a phone with stock firmware
and factory-restored settings and data. 

Each of the four payloads succeeded in compromising the phones security
or data integrity. A list of the tested payloads is as follows:
\begin{itemize}
\item Disable all user capability checks in setuid() and setgid() system
calls. This allows any user and app to achieve root privileges with
a simple system-call.
\item Silently incapacitate the Security Enhanced Linux (SELinux) \cite{smalley2001implementing}
module. While SELinux will still report blocking suspicious activity,
it will not actually be blocked.
\item Create a user exploitable system-wide vulnerability. The buffer check
is disabled for all user buffers on system calls, resulting in many
different vulnerabilities exploitable through many techniques.
\item Create a hidden vulnerability within the kernel. A specific kernel
vulnerability is generated, functioning as a backdoor for a knowledgeable
attacker while remaining hidden.
\end{itemize}

%% file: end-to-end-attack.tex
\begin{table*}
\begin{centering}
\caption{\label{tab:touch-based-time}A summary of our demonstrated attacks.}
\small%
\begin{tabular}{|c|c|c|c|}
\hline 
Attack & Time to execute & Screen Blanked? & Video Demo\tabularnewline
\hline 
\hline 
Malicious Software Installation & 21 seconds & Yes & \url{https://youtu.be/83VMVrcEOCM}\tabularnewline
\hline 
Take Picture and Send Via Email & 14 seconds & Yes & \url{https://youtu.be/WS4NChPjaaY}\tabularnewline
\hline 
Replace URL with phishing URL & <1 second & No & \url{https://youtu.be/XZujd42eYek}\tabularnewline
\hline 
Log and exfiltrate screen unlock pattern & 16 seconds & Yes & \url{https://youtu.be/fY58zoadqMA}\tabularnewline
\hline 
Complete Phone Compromise & 65 seconds & Yes & \url{https://youtu.be/sDfD5fJfiNc}\tabularnewline
\hline 
\end{tabular}\normalsize
\par\end{centering}
\end{table*}

While each of the attacks described in Section \ref{sec:Attack-Building-Blocks}
poses a threat on their own, a combination of these attacks can lead
to an even more powerful attack. We summarize the attacks presented
in this Section in Table \ref{tab:touch-based-time}, complete with
demonstration videos, and describe each of the attacks below.

\subsection{User Impersonation and User Compromise}

The basis for the user impersonation and compromise parts of this
attack are the touch logging and injection capabilities described
in Subsection \ref{subsec:Touch-Logging-and}. These capabilities
can be extended and used for a variety of malicious activities. Since
our attack model assumes a malicious touchscreen assembly, the attacker
can turn off power to the display panel while a malicious action is
performed, allowing most attacks to be carried out stealthily.

The first attack we demonstrate is the \textbf{malicious software
installation} attack. As illustrated in the video, this attack installs
and starts an app from the Google Play Store. By using Android's internal
search functionality, the attacker can type in the name of the Play
Store app instead of searching for it onscreen, making our attack
more resilient to users who customize their home screens. It is important
to note that the attack can install an app with arbitrary rights and
permissions, since the malicious touchscreen can confirm any secrity
prompt posed by the operating system. This attack takes less than
30 seconds, and can be performed when the phone is unattended and
when the screen is powered off.

Next, we show how the malicious touchscreen can \textbf{take a picture
of the phone's owner and send it} to the attacker via email. As seen
in the video, this attack activates the camera and sends a 'selfie'
to the attacker. This attack also takes less than 30 seconds, and
can be performed while the display is turned off, allowing the attack
to be carried out without the user's knowledge.

Our third attack shows how the malicious screen can stealthily \textbf{replace
a hand-typed URL with a phishing URL}. As the video shows, this attack
waits for the user to type a URL, then quickly replaces it with a
matching phishing URL. The confused user can then be enticed to type
in his or her credentials, assuming that a hand-typed URL is always
secure. This attack takes less than 1 second, but uniquely requires
the screen to be turned on and the user present, thus risking discovery.
We note that our current attack setup has a typing rate of over 60
characters per second.

Our fourth attack shows how the malicious screen can \textbf{log and
exfiltrate the user's screen unlock pattern }to an online whiteboard
website. The video demonstrates how the attack records the user's
unlock pattern and draws it over a shared online whiteboard, which
is shared via the Internet with the attacker's PC. This attack demonstrates
both the collection and the infiltration abilities of the attack vector.
This attack also takes less than 30 seconds, and its exfiltration
step can also be performed while the screen is turned off.

Our final attack completely compromises the phone, disables SELinux,
and opens a reverse shell connected to a remote attacker. This attack
is unique in that it requires an exploitable bug in the third-party
device driver code. We describe this attack in more detail in the
following Subsection.

\subsection{Phone Compromise}

\begin{figure*}
\begin{centering}
\includegraphics[width=0.6\textwidth]{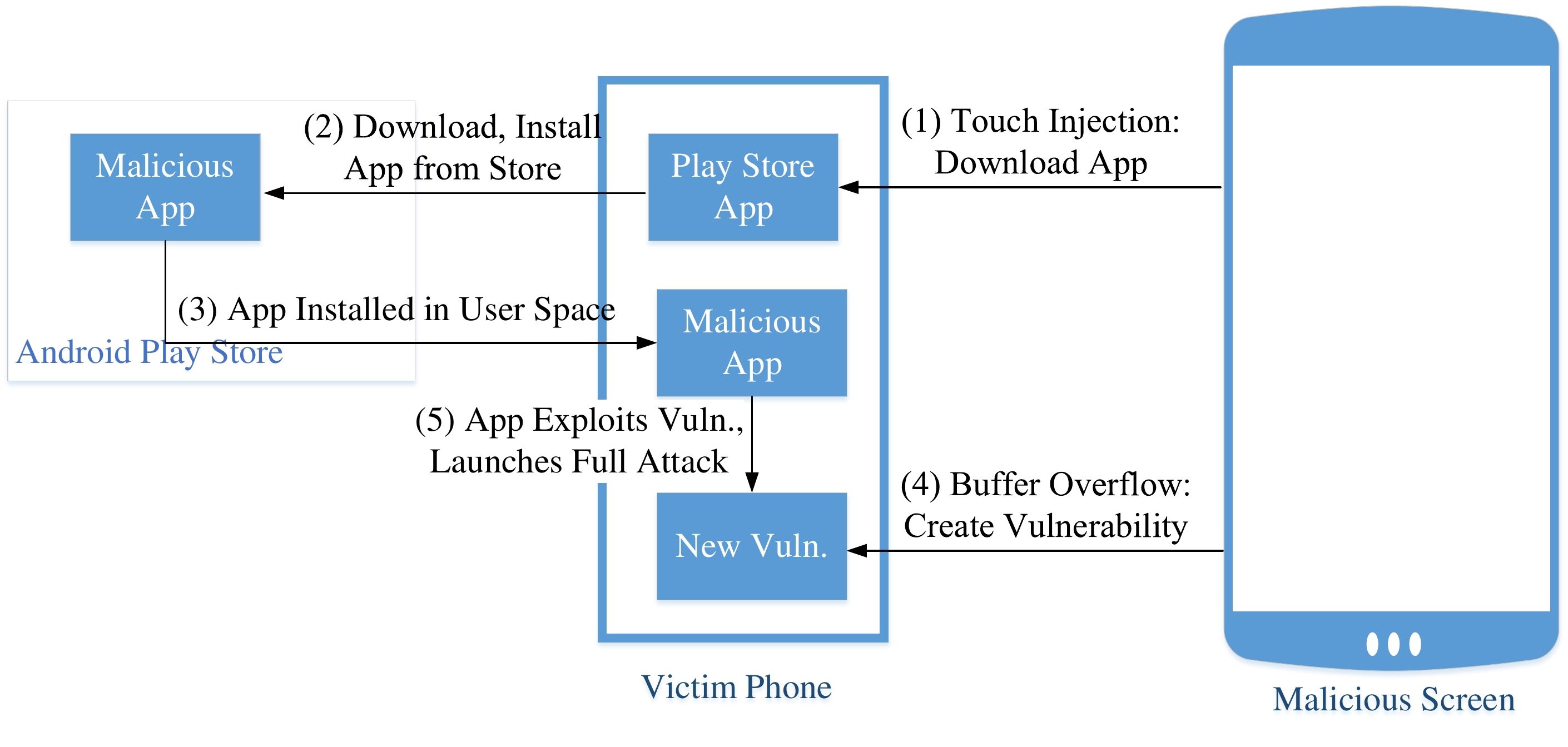}
\par\end{centering}
\caption{\label{fig:Fully-compromising-the}Fully compromising the phone using
a malicious touchscreen.}
\end{figure*}

To completely compromise the phone, we use a combination of touch
events and driver exploits, as illustrated in Figure \ref{fig:Fully-compromising-the}:
First, 

the attacker uses \textbf{touch injection} to install an innocent-looking
app from the Google Play app market. The next time the phone restarts,
the malicious micro-controller initiates \textbf{kernel exploitation
}during the boot sequence and creates a vulnerability in the kernel
that is exploitable by user app. Once the phone completes booting,
the previously installed app uses the vulnerability created by the
micro-controller to take control of the system and perform malicious
activity. The malicious app then reboots the phone and the now-compromised
phone resumes normal activity.

\subsubsection{Implementation}

For this demonstration, a user app was created and uploaded to the
Google Play app market. The app starts when the phone boots up and
performs a series of system calls by writing to the pesudo-file ``/prof/self/clear\_refs''.
While the phone is in a normal state, these system calls cause no
issues and should raise no suspicion. During the exploitation of the
kernel by the malicious micro-controller, the actions of the pesudo-file
``/prof/self/clear\_refs'' are modified, and a vulnerability is
introduced to it. This causes a change in the behavior of the app
which is now able to exploit that vulnerability and execute code in
kernel context. We note that since the app is designed to exploit
a vulnerability that is non-existent under normal conditions, it appears
completely benign when a malicious screen is not present. This enabled
our app to overcome malware filters and detectors, including Google
Play's gatekeeper, Google Bouncer.

Once the app has gained the ability of executing commands with kernel
permissions, it elevates privileges to root, deactivates the SELinux
protection module, exfiltrates application private data and authentication
tokens and submits the data to an online server, and finally creates
of a root shell enabling an attacker to gain remote access. A video
demonstration of the attack is available at \url{https://youtu.be/Z_esD1Z78Ms}

%% file: additional-devices.tex
While the main attack demonstrated here is crafted to the Nexus 6P
phone, many more phones use similar device drivers \cite{ourPaper}.
A small scale review performed by the authors on three additional
phones that contain a Synaptics touch controller (Samsung Galaxy S5,
LG Nexus 5x, LG Nexus 5) shows similar vulnerabilities to the ones
exploited in the attack described here. 

To further demonstrate the generality of our attack method, we extended
it to another target device with a different hardware architecture.
The device we investigated as an LG G Pad 7.0 (model v400) tablet.
This devce runs the Android 5.0.2 operating system and contains a
different touchscreen controller than the Nexus 6P phone.

\subsection{Tablet Hardware}

\begin{figure}
\begin{centering}
\includegraphics[width=0.95\columnwidth]{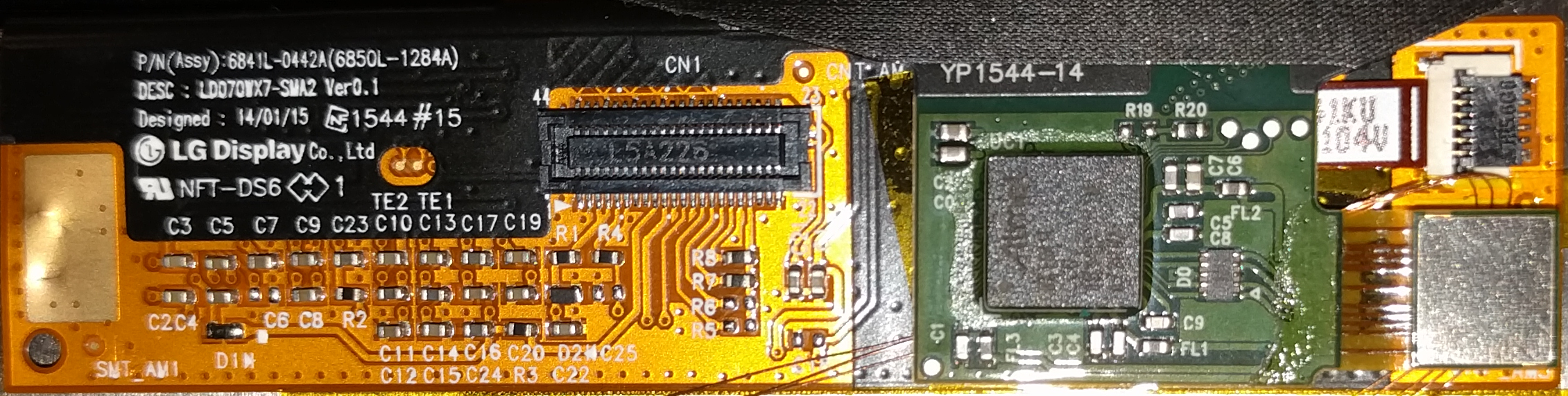}
\par\end{centering}
\caption{\label{fig:The-LG-v400}The LG v400 display assembly PCBs, showing
the motherboard with a flex cable connector and an Atmel T641 touchscreen
controller mounted on a daughter board. Also visible: 0.1mm thick
wires soldered to the connector between the boards.}

\end{figure}

The LG v400 tablet employs an Atmel T641 touchscreen controller. The
screen assembly is designed similarly to other devices such as the
Nexus 6P and is built of a main display motherboard and a smaller
touch controller daughter board. The display assembly boards are attached
with Board-To-Board connectors and can be separated for maintenance.
The PCBs belonging to the display module can be seen in Figure \ref{fig:The-LG-v400}.

\subsection{Similarities to the Synaptics Touchscreen Controller}

While the touchscreen controllers of the LG v400 and Nexus 6P were
designed and produced by different vendors, there are shared similarities
among the controllers and their device drivers. In the hardware aspect
it is notable that both controllers communicate via an I$^{2}$C bus
in 400 kHz \emph{Fast-mode} and both controllers signal of incoming
events using a designated interrupt line. The protocols used by both
controllers utilize an entity-based framework where specific controller
functionalities are accessed via an entity. The information about
entities contained within the controller is retrieved on every phone
boot from the touchscreen controller by the device driver.

\subsection{Attaching the Malicious Hardware}

Enameled copper wires were soldered to the Board-To-Board connector
of the display assembly motherboard, a Saleae logic analyzer was connected
to the wires and normal touchscreen controller behavior was recorded.

\subsection{Attacking the device driver}

An STM32L432 micro-controller module was connected to the communication
lines belonging to the touch controller and the original touch controller
daughter board was disconnected. The micro-controller was programmed
to replay previously recorded responses of a genuine touch controller.
Inspection of the device driver revealed unsafe buffer handling in
numerous locations. By falsely reporting an irregularly large entity,
the malicious micro-controller was able to cause the device driver
to read 2048 bytes from the bus into an 80-byte global array. The
buffer overflow affected kernel memory and resulted in the overrun
of various internal pointers and eventually a crash.

While the attack shown in this section is not complete, these preliminary
results show how the complete attacks shown in sections \ref{sec:Attack-Building-Blocks}
and \ref{sec:End-to-End-Attacks} can be implemented on additional
devices with different peripherals. 

In addition, the similarity in different peripheral implementation
makes adapting existing attacks to new peripherals easier. For example,
after reverse engineering the touch reporting mechanism of the Atmel
touch controller, the Synaptics touch injection attack can be copied
over to devices with an Atmel touch controller, even without discovering
any vulnerability in the Atmel device driver.

%% file: discussion.tex
\subsection{Toward Low-Cost Active Fault Attacks}

Many studies have tried to compromise the integrity of code running
on the secure system's CPU via \textbf{software-oriented attack} methods
(such as buffer overflows, return-oriented programming and so on).
The advantage of a software-oriented attack is its ease of execution
\textendash{} an attacker does not need physical access to the device
under attack, only code execution privileges, making it possible to
mount attacks remotely and at scale. However, the widespread prevalence
of software-based attack methods lead to a serious effort to protect
CPUs from this direction of attack using countermeasures such as sandboxing,
user and root privilege separation, and hardware-assisted trusted
execution environments. Despite the original delivery vector, our
attack is still software oriented in nature.

On the other extreme of the attack spectrum, studies have also attempted
to use hardware-based \textbf{active fault attacks} to compromise
the main CPU's integrity using invasive methods such as laser fault
injection, FIB-based circuit editing and side-channel attacks. These
attacks can effectively deal with software-based countermeasures,
for example by disabling various security-oriented parts of the secure
device or by exposing additional sources of secret information that
can assist in device compromise. The downside of such an attack is
its high cost and effort for the attacker \textendash{} in most cases,
these attacks require that the attacker have complete physical control
over the device under attack, and that the attacker furthermore has
a considerable degree of budget and technical expertise. This makes
the threat of active fault attacks less relevant in many attack models. 

The concept of attacking secure devices via \textbf{malicious replacement
units }may allow an interesting trade-off between the two methods
of software-oriented attacks and active fault attacks. This is because
it provides an attacker with a low-risk method of getting ``up close
and personal'' to the main CPU's hardware interfaces, while at the
same time requiring very little of the attacker in terms of attack
cost or time spent. This, in turn, makes it possible to carry out
active fault attacks without a dedicated effort from the attacker.
Moreover, compromise of such a device might be done in a way which
cannot be detected by the main CPU by leaving no software traces.

\subsection{The Case for Hardware-Based Countermeasures}

The unique attack model we discuss in our paper allows us to ``fight
hardware with hardware''. In order to protect the phone from a malicious
replacement screen, we propose implementing a low-cost, hardware-based
solution in the form of I$^{2}$C interface proxy firewall. Such a
firewall can monitor the communication of the I$^{2}$C interfaces
and protect the device from attacks originating from the malicious
screen. Placing this device on the motherboard means that it will
not be affected by malicious component replacement. The use of a hardware
countermeasure allows for protection against both added malicious
components and modified firmware attacks. It may also detect malicious
behavior of firmware code that was modified by an insider and may
be officially signed or encrypted. Since it does not require any changes
on the CPU or component side, this solution should be much faster
to implement than cryptographically-based approaches such as I$^{2}$C
encryption or device authentication.

\subsection{Responsible Disclosure}

The authors followed responsible disclosure practices by disclosing
the Synaptics device driver vulnerabilities to Google on Feb. 16,
2017. The disclosure includes the details necessary for understanding,
reproducing, and fixing of the issues discovered during the research.
Google acknowledged the reported issues and issued a CVE (CVE-2017-0650)
with critical severity.

The vulnerabilities discovered in the Atmel device driver are being
compiled into a responsible disclosure report at the time of submitting
this paper.

\subsection{Future Work}

While this paper shows critical issues with smartphone software and
hardware infrastructure, it mainly focuses on one phone model. Performing
a wider analysis on multiple phone models and peripherals will help
understand how vulnerable are the majority of phones used worldwide. 

A root-cause analysis on the vulnerabilities found can shed light
on which of the vendor's design and implementation processes contributed
to the forming of such vulnerabilities. Such insights can help in
development of techniques for design flaw mitigation and might yield
recommendations for efficient and secure design of hardware and software
elements. Additional techniques can be attempted for exploitation
by malicious peripherals such as replacing the firmware in an embedded
component and creating an attack without the use of external components.

\subsection{Conclusions}

The threat of a malicious peripheral existing inside consumer electronics
should not be taken lightly. As this paper shows, attacks by malicious
peripherals are feasible, scalable, and invisible to most detection
techniques. A well motivated adversary may be fully capable of mounting
such attacks in a large scale or against specific targets. System
designers should consider replacement components to be \emph{outside}
the phone's trust boundary, and design their defenses accordingly.

%% file: smashing-woot.bbl
\begin{thebibliography}{10}

\bibitem{cracked}
\relax Motorola~Mobility.
\newblock Cracked screens and broken hearts - the 2015 motorola global
  shattered screen survey.
\newblock
  \url{https://community.motorola.com/blog/cracked-screens-and-broken-hearts}.

\bibitem{disaApproved}
\relax Defense Information Systems~Agency.
\newblock {\em The Department of Denfense Approved Products List}.
\newblock \url{https://aplits.disa.mil/processAPList}.

\bibitem{zhou2012dissecting}
Yajin Zhou and Xuxian Jiang.
\newblock Dissecting android malware: Characterization and evolution.
\newblock In {\em 2012 IEEE Symposium on Security and Privacy}, pages 95--109.
  IEEE, 2012.

\bibitem{bickford2010rootkits}
Jeffrey Bickford, Ryan O'Hare, Arati Baliga, Vinod Ganapathy, and Liviu Iftode.
\newblock Rootkits on smart phones: attacks, implications and opportunities.
\newblock In {\em Proceedings of the eleventh workshop on mobile computing
  systems \& applications}, pages 49--54. ACM, 2010.

\bibitem{corbet2005linux}
Jonathan Corbet, Alessandro Rubini, and Greg Kroah-Hartman.
\newblock {\em Linux Device Drivers: Where the Kernel Meets the Hardware}.
\newblock " O'Reilly Media, Inc.", 2005.

\bibitem{danisevskis2013dark}
Janis Danisevskis, Marta Piekarska, and Jean-Pierre Seifert.
\newblock Dark side of the shader: Mobile gpu-aided malware delivery.
\newblock In {\em International Conference on Information Security and
  Cryptology}, pages 483--495. Springer, 2013.

\bibitem{ladakis2013you}
Evangelos Ladakis, Lazaros Koromilas, Giorgos Vasiliadis, Michalis
  Polychronakis, and Sotiris Ioannidis.
\newblock You can type, but you can't hide: A stealthy gpu-based keylogger.
\newblock In {\em Proceedings of the 6th European Workshop on System Security
  (EuroSec)}, 2013.

\bibitem{brocker2014iseeyou}
Matthew Brocker and Stephen Checkoway.
\newblock iseeyou: Disabling the macbook webcam indicator led.
\newblock In {\em USENIX Security}, pages 337--352, 2014.

\bibitem{appleError53}
\relax Apple~Inc.
\newblock {\em Error 53 support page}.
\newblock \url{https://support.apple.com/en-il/HT205628}.

\bibitem{nistDraft}
\relax National Institute~of Standards and Technology.
\newblock {\em Cybersecurity Framework v1.1 - Draft}.
\newblock \url{https://www.nist.gov/cyberframework/draft-version-11}.

\bibitem{zhou2014peril}
Xiaoyong Zhou, Yeonjoon Lee, Nan Zhang, Muhammad Naveed, and XiaoFeng Wang.
\newblock The peril of fragmentation: Security hazards in android device driver
  customizations.
\newblock In {\em 2014 IEEE Symposium on Security and Privacy}, pages 409--423.
  IEEE, 2014.

\bibitem{ourPaper}
Omer Shwartz, Guy Shitrit, Asaf Shabtai, and Yossi Oren.
\newblock From smashed screens to smashed stacks: Attacking mobile phones using
  malicious aftermarket parts.
\newblock In {\em Workshop on Security for Embedded and Mobile Systems (SEMS
  2017)}, 2017.

\bibitem{secbull}
\relax Google.
\newblock {\em Android Security Bulletin}.

\bibitem{Counterfeit}
\relax UL.
\newblock Counterfeit iphone adapters.

\bibitem{CounterfeitApple}
Amit Chowdhry.
\newblock Apple: Nearly 90
  counterfeit.
\newblock {\em Forbes.com}, 2016.

\bibitem{butt2009protecting}
Shakeel Butt, Vinod Ganapathy, Michael~M Swift, and Chih-Cheng Chang.
\newblock Protecting commodity operating system kernels from vulnerable device
  drivers.
\newblock In {\em Computer Security Applications Conference, 2009. ACSAC'09.
  Annual}, pages 301--310. IEEE, 2009.

\bibitem{okolie2013penetration}
CC~Okolie, FA~Oladeji, BC~Benjamin, HA~Alakiri, and O~Olisa.
\newblock Penetration testing for android smartphones.
\newblock 2013.

\bibitem{guri2015joker}
Mordechai Guri, Yuri Poliak, Bracha Shapira, and Yuval Elovici.
\newblock Joker: Trusted detection of kernel rootkits in android devices via
  jtag interface.
\newblock In {\em Trustcom/BigDataSE/ISPA, 2015 IEEE}, volume~1, pages 65--73.
  IEEE, 2015.

\bibitem{suarez2014evolution}
Guillermo Suarez-Tangil, Juan~E Tapiador, Pedro Peris-Lopez, and Arturo
  Ribagorda.
\newblock Evolution, detection and analysis of malware for smart devices.
\newblock {\em IEEE Communications Surveys \& Tutorials}, 16(2):961--987, 2014.

\bibitem{msmkernel}
\relax Google.
\newblock {\em Android Kernel tree for Qualcomm chipsets}.
\newblock \url{https://android.googlesource.com/kernel/msm.git}.

\bibitem{NXP2014i2c}
\relax NXP.
\newblock {\em I2C-bus specification and user manual}, April 2014.
\newblock \url{http://www.nxp.com/documents/user\_manual/UM10204.pdf}.

\bibitem{arduino}
\relax Arduino.
\newblock {\em Arduino Home Page}.
\newblock \url{https://www.arduino.cc}.

\bibitem{atmega}
\relax Atmel~Corporation.
\newblock {\em ATmega Datasheet}, November 2015.
\newblock
  \url{http://www.atmel.com/images/atmel-8271-8-bit-avr-microcontroller-atmega48a-48pa-88a-88pa-168a-168pa-328-328p_datasheet_summary.pdf}.

\bibitem{cypress}
\relax Cypress.
\newblock {\em PSoC 4000 Family Datasheet}, November 2017.
\newblock \url{http://www.cypress.com/file/138646/download}.

\bibitem{DBLP:journals/tifs/FrankBMMS13}
Mario Frank, Ralf Biedert, Eugene Ma, Ivan Martinovic, and Dawn Song.
\newblock Touchalytics: On the applicability of touchscreen input as a
  behavioral biometric for continuous authentication.
\newblock {\em {IEEE} Trans. Information Forensics and Security},
  8(1):136--148, 2013.

\bibitem{hund2009return}
Ralf Hund, Thorsten Holz, and Felix~C Freiling.
\newblock Return-oriented rootkits: Bypassing kernel code integrity protection
  mechanisms.
\newblock In {\em USENIX Security Symposium}, pages 383--398, 2009.

\bibitem{smalley2001implementing}
Stephen Smalley, Chris Vance, and Wayne Salamon.
\newblock Implementing selinux as a linux security module.
\newblock {\em NAI Labs Report}, 1(43):139, 2001.

\end{thebibliography}
